\documentclass[showkeys,showpacs,prl,onecolumn]{revtex4-2}
\usepackage{graphicx}
\usepackage{dcolumn}
\usepackage{bm}
\usepackage{verbatim}
\usepackage{multirow}
\usepackage{amsmath}
\usepackage{amssymb}
\usepackage{float}
\usepackage{mathrsfs}

\newcommand{\cmmnt}[1]{}

\begin{document}
\preprint{ORNL}

\title{Ancillary  entangling Floquet kicks for accelerating quantum algorithms}

\author{C.-C. Joseph Wang, Phillip C. Lotshaw, Titus Morris, Vicente Leyton-Ortega, Daniel Claudino, and Travis S. Humble}
 \affiliation{Quantum Computational Science Group, Quantum Information Science Section, Oak Ridge National Laboratory, 1 Bethel Valley Road, Oak Ridge, 37831, TN, USA}
\email[]{wangccj@ornl.gov}
\thanks{This manuscript has been authored by UT-Battelle, LLC, under Contract No. DE-AC0500OR22725 with the U.S. Department of Energy. The United States Government retains and the publisher, by accepting the article for publication, acknowledges that the United States Government retains a non-exclusive, paid-up, irrevocable, worldwide license to publish or reproduce the published form of this manuscript or allow others to do so, for the United States Government purposes. The Department of Energy will provide public access to these results of federally sponsored research by the DOE Public Access Plan.}

\date{\today}
\begin{abstract} 
Quantum simulation with adiabatic annealing can provide insight into difficult problems that are impossible to study with classical computers. However, it deteriorates when the systems scale up due to the shrinkage of the excitation gap and thus places an annealing rate bottleneck for high success probability.  
Here, we accelerate quantum simulation using digital multi-qubit gates that entangle primary system qubits with the ancillary qubits.
The practical benefits originate from tuning the ancillary gauge degrees of freedom to enhance the quantum algorithm's original functionality in the system subspace. For simple but nontrivial short-ranged, infinite long-ranged transverse-field Ising models, and the hydrogen molecule model after qubit encoding, we show improvement in the time to solution by one hundred percent but with higher accuracy through exact state-vector numerical simulation in a digital-analog setting. The findings are further supported by time-averaged Hamiltonian theory. 
\end{abstract}


\maketitle
Classical simulation of quantum many-body systems is incredibly difficult because of the many degrees of freedom (DOF) involved~\cite{QS}. To solve this problem, quantum analog simulation with adiabatic annealing has been used, which requires less strict controls for implementation~\cite{QA1,QA2,QA3}. However, due to the critical slowdown caused by the decrease in the minimal excitation gap as the quantum system becomes larger~\cite{QCP}, the initial success of adiabatic quantum annealing protocols in small quantum systems~\cite{CM1, CM2} is not expected to continue at a larger scale
due to the upper bound of the coherence time set by quantum hardware and the condition to stay adiabatic. Digital quantum simulations~\cite{Seth} have become the mainstream
to sidestep the limitation. 

More recently, 
digital simulations through hybrid variational quantum algorithms (HVQAs) with classical optimization show great promise for realizing applications~\cite{reviewVQA1,cerezo2021variational,xu2021variational,Qchem1,Qchem2,Qchem3,biamonte2021universal} due to their universality (in computation) and the potential to demonstrate quantum advantage in practical tasks. However, with overparametrized circuits, HVQAs also suffer from many problems inherited from classical optimization~\cite{Review_Barren_Plateaus,Barren,Cerezo,VQA_PROBLEM,VQA_NP_HARD}. 
To leverage practical values from current noisy quantum computing hardware~\cite{preskill2018quantum} toward quantum utility, diverse universal quantum computing paradigms---involving the programmable analog model, gate models, digital-analog models, and measurement-based feedforward model---are actively pursued~\cite{Altman,Quantinuum,DWAVE,QUERA,IBM,IONQ,PSIQ,PRXQuantum.2.020328}. The race to design practical quantum algorithms (heuristic or deterministic) compatible with diverse paradigms and platforms for realizing practical quantum applications at scale remains critical. We show that quantum simulations can be accelerated with optimal quantum kicks, requiring only single parameter optimization offline (insensitive to system size) and utilizing limited resources of ancillary qubits while preserving the quantum information in the system qubits where the core quantum application operates.

We adopt a quantum annealer as the quantum optimizer with encoded applications in the system-only Hamiltonian.
The quantum annealer can overcome the local minima traps by exploring the system energy landscape with mixer system Hamiltonian through quantum tunneling~\cite{RevModPhys.80.1061}. It is theoretically equivalent to the quantum approximate optimization algorithm after dense Trotterization in a finite simulation time. However, the annealer faces a significant drawback due to the finite excitation gap of the encoded Hamiltonian not being large enough for the annealing time scale. Thus the quantum dynamics cannot follow the desired adiabatic state evolution for the system Hamiltonian at a faster annealing.
Although counter-diabatic Hamiltonian approaches have been proposed to counter the diabatic effects, the complexity of the nested counter-diabatic terms makes these approaches hard to implement in practice~\cite{PhysRevLett.123.090602,counter-diabatic,Wurtz2022counterdiabaticity}. 
To address this issue, we propose a new quantum protocol to accelerate quantum simulation by controlling nonadiabatic quantum state evolution with additional ancillary qubits. We achieve this by designing entangling gates with periodic (Floquet) kicks that deviate state evolution from the adiabatic path early in the simulation but allow the passage back to the target quantum path shortly after. 

First, we enlarge the primary system Hilbert space $S$ with the secondary ancillary Hilbert space $A$.
The quantum computation is enabled by the always-on {\it system} unitary $U^S(t)$ that encodes the interested problem in analog quantum simulation and the digital quantum kick unitaries $U^{SA}({\hat \Theta}^{A}, \hat S)=\exp(-i{\hat \Theta}^{A} \otimes \hat S)$ that accelerate the quantum simulation. The
operator $\hat \Theta^{A}$ and the collective system operator $\hat S = \sum_{l=1}^{N_S}\sigma_{l}^{S}$ (with $N_{S}$ system qubits) act on the $A$ space and $S$ space respectively.
The choice of $\hat \Theta^{A}$ is not unique but can be limited to simple forms with minimal compilation for target quantum simulation. We initialize the ancillary state as a product state right at the outset for the quantum simulation before the entanglement with the system qubits enabled by $U^{SA}(t)$. In qubit language, the corresponding operator ${\hat {\Theta}}^{A}$ is given by
\begin{equation}
    {\hat {\Theta}}^{A} = \sum_{l=1}^{N_{A}} \theta_{l} \sigma_{l}^{A}
    \label{EQ:1}
\end{equation}
where the Pauli operators $\sigma_{l}^{A}$ for qubit $l$ are selected from the gate set $\{X_{l}^{A}, Y_{l}^{A}, Z_{l}^{A}\}$ and the summation is over $N_A$ ancillary qubits.
For ground-state properties,
the target function to minimize is the reduced system energy in $S: Tr_{S}(\hat{\rho}^{S}(t)H^{S}(t))$ after the repetitive action of the identical kick unitary operator
$U^{SA}(\hat \Theta^{A})$ acting on the state ${\hat \rho}^{SA}(t-\epsilon)|_{\epsilon \rightarrow 0^{+}} := |\psi^{SA}(t-\epsilon)\rangle\langle
\psi^{SA}(t-\epsilon)|$ governed by the always-on system Hamiltonian right before the quantum kick. 
The overall quantum state evolves as ${\hat \rho}^{SA}(t) = U^{\dagger SA}(t) \rho^{SA}(t-\epsilon)U^{SA}(t)$ 
with the reduced system state after tracing out ancillary subsystem as ${\hat \rho}^{S}(t) = Tr_{A}{\hat \rho}^{SA}(t)$.
To have a low-cost compilation, we set our smart gate parameters $\theta_{l}$ to be a single uniform parameter $\theta$ for the simplest gate design with one spin orientation selected from the ancillary Pauli operator set $\{X_{l}^{A}, Y_{l}^{A}, Z_{l}^{A}\}$.

Secondly, the choices of efficient multi-qubit quantum kicks---up to a local unitary rotation in {\it A} (gauge DOF)--- depend on the encoded quantum simulation applications.
The quantum kick gate sets shall be selected as those   
with good sensitivity to change the reduced system energy 
landscape $Tr_{S}\left({\hat\rho}^{S} H^{S}(t)\right)$ 
and meet the non-commuting relation $[\sigma_{i}^{S}, H^{S}(t=0)] \neq 0$ and the commutation relation $[\sigma_{i}^{S}, H^{S}(t=T)]=0$ at the end of simulation time $T$.
Guided by numerical studies from our in-house state-vector simulator and theoretical studies with time-averaged Hamiltonian theory~\cite{Intro_averaged_H,time-averaged-H, Vicente}---based on leading-order Magnus expansion in the strong rotating reference frame from the trivial transverse-field (mixer) Hamiltonian.
In digital-analog settings, we identify that discrete entangling quantum kicks are instrumental in accelerating applications when the quantum kick strength is optimally upper-bounded without causing nonadiabatic excitations.

{\it Optimal kick angle estimation.---}
We theoretically estimate the optimal kick angle $\theta_{opt}$ based on the non-perturbative time-averaged Hamiltonian approach. The theory estimation reconciles with the numerical results from the state-vector simulation.
The optimal kick angle to stay close to the system ground state energy
for the quantum kicks, is given by the following expression (See details in~\cite{SM0}) for the TFIM models and the $\rm H_2$ molecule model for describing quantum dynamics before dynamical quantum phase transition (QPT) with a theory cutoff $T\approx\tau$:
\begin{equation}
\theta_{opt} \approx \frac{1}{N_A N_K}\sqrt{1-E_{S}^{T}/E_{S}^{Mixer}(\tau)},
\label{EQ:3}
\end{equation}
in which $E_{S}^{T}$ is the true system ground state energy and $E_{S}^{Mixer}(T=\tau)$ is the system mixer energy. The formula is 
Based on the numerical observation for circumstances with over-kicking angles that cause reduced system energy overshoot above true system ground state energy before QPT, it never ends up closer to the true ground state energy.
The expression is valid and universal as long as the mixer Hamiltonian and the kick Hamiltonian are the {\it only} dominant~\cite{SM0} Hamiltonian for the proposed quantum kick algorithm since the specifics of the system Hamiltonian will not contribute much to quantum state evolution before QPT. As follows, we will validate the ideas through quantum Ising models universal for simulating exotic
quantum phases~\cite{IS} and the hydrogen molecule model, the fundamental building block for quantum chemistry simulation. 

{\it Nearest-neighbor transverse field Ising model (NN-TFIM) in an open chain.---}
It is well understood that a QPT occurs for the NN-TFIM~\cite{sachdev_2011} with a finite excitation gap $\Delta_{min}$ that sets the bound
to the simulation time to solution.
With the kick protocol, we can speed up the passage through the QPT even at a faster annealing rate than $\Delta_{min}^{-1}$. 
The schedule for the system Hamiltonian $H^{S}(t)$ with the end simulation time $T$ is given by:
\begin{equation}
    H^{S}(t) = \underbrace{\sum\limits_{i}\theta_{M}(t)Z^{S}_{i}}_{H^{S}_{M}(t)} + \underbrace {\sum\limits_{\left< i < j \right >}-\theta^{0}_{XX}X^{S}_{i}X^{S}_{j}}_{H^{S}_{P}}, 
    \label{EQ:2}
\end{equation}
in which $\theta^{0}_{XX} > 0$ and $H^{S}_M$ is the one-local mixer Hamiltonian in $S$ space with the positive amplitude $\theta_{M}(t) = \theta_{Z}^{0}(:=\theta_{M}(0)) e^{-t/\tau}$
following the spatially uniform annealing with the time constant $\tau$. 
The always-on negative uniform Ising interaction $-\theta^{0}_{XX}$ between system qubits
favors the \textrm{GHZ} ground state $|\textrm{GHZ}\rangle=1/\sqrt{2}(|0_{x}{\rangle}^{\otimes N_S} + |1_{x}{\rangle}^{\otimes N_S})$ for the system problem Hamiltonian $H^{S}_{P}$. With the chosen parameters $\theta_{Z}^{0}/\theta^{0}_{XX} = 5.0$, the model is initialized
at the ground state in proximity to the product state $|1\rangle^{\otimes N_S}$ for the system qubits and $|0\rangle^{\otimes N_A}$ for the ancillary qubits.

The summation of the probability measured at the computational basis $|0\rangle^{\otimes N_S}$ and $|1\rangle^{\otimes N_S}$ is provided as the proxy for the ferromagnetic order (polarized state) in the basis and abbreviated as PFM. We expect that, with faster annealing ($\Delta_{min}\tau > 1$),
diabatic transitions are not negligible and limit the time scale $\tau$ to reach the true target ground state.
This is the critical slowdown effect. However, this is not the problem with a large transverse field at the initial stage of the quantum simulation with a large excitation gap proportional to $\theta_{Z}^{0}$.

\begin{figure}[h]
    \centering
    \includegraphics[scale=0.60]{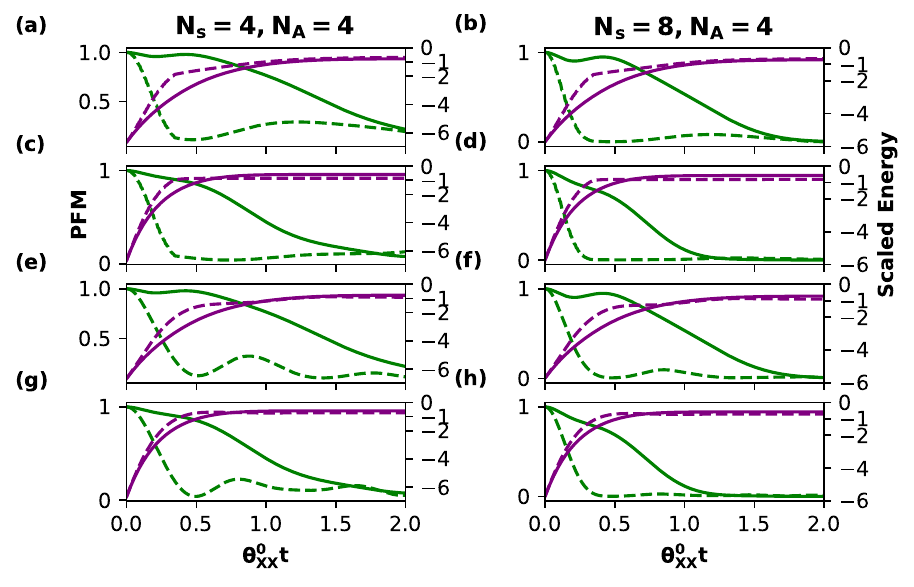}
    \caption{NN-TFIM in an open chain with $\theta_{XX}^{0} = 1.0,~\theta_{M}(0) = \theta_Z^{0} = 5\theta_{XX}^{0}$: time dependency of the projected probability PFM in the ground state for the mixer Hamiltonian $|1\rangle^{\otimes N_S}$ (in green color with values given by the left side ticks) and the scaled system energy (in purple with values given by the right side ticks) computed by the ratio between the reduced system energy and the absolute value of the true system ground-state energy. 
    The purple solid lines present the cases without the kicks $\theta = 0$. The green lines (dashed ones with kicks and solid ones without kicks) are the proxy for the disappearance of the ferromagnetic order along the Pauli-$X$ orientation. 
    All cases with solid (dashed) lines represent the cases without (with) any types of kicks. 
    Floquet kicks with the same kick angle $\theta = 0.004$ and the kick time interval $\Delta t_{K}$  between kicks is related to the Ising interaction $\theta_{XX}^{0}\Delta t_{K} = 0.0008$. (a) $\theta_{Z}^{0}\tau = 0.8^{-1}$, (b) $\theta_{Z}^{0}\tau = 0.8^{-1}$, (c) $\theta_{Z}^{0}\tau = 1.6^{-1}$, (d) $\theta_{Z}^{0}\tau = 1.6^{-1}$. Continuous weak kicks with the constant kick angle $\theta \rightarrow 0, N_{K}\rightarrow \infty$: (e) $\theta_{Z}^{0}\tau = 0.8^{-1}$, (f) $\theta_{Z}^{0}\tau = 0.8^{-1}$, (g) $\theta_{Z}^{0}\tau = 1.6^{-1}$, (h) $\theta_{Z}^{0}\tau = 1.6^{-1}$.}
    \label{FIG:2}
\end{figure}

To speed up the simulation, we can apply an efficient quantum kick that does not commute with the 
trivial mixer Hamiltonian $H^{S}_{M}(t)$ at the kick time with a properly chosen ancillary and system Pauli operators $\sigma_{l}^{A}$ and $\sigma_{l}^{S}$ at this early stage in which the diabatic effects are suppressed mostly.
In Fig.~\ref{FIG:2}, we show the efficient digital kick Hamiltonian $H^{SA}(t) = \theta(t)\sum_{ij} X^{S}_{i} \otimes Z^{A}_{j}$ at time $t$ with Dirac kicks $\theta(t):= \theta \sum_{K=0}^{N_{K}-1} \delta(t-K\Delta t_{K})$ in which $N_{K}$ is the accumulated number of kicks before time $t$. The impulsive kicks
intercept periodically with the always-on system Hamiltonian ($H_{P}^{S} + H_{M}^{S}$) with a given kick angle $\theta$ immediately after the state preparation. In the numerical simulation, the optimal kick angle $\theta_{opt}$ is found empirically by
grid searched results, and is found later in excellent agreement~\cite{SM0} with the theory estimation from the time-averaged Hamiltonian~\cite{Intro_averaged_H,time-averaged-H}. 
We differentiate the efficiency of kicks by the sensitivity response of the system to each type of kicks~\cite{SM1}. 
The system Pauli operator $X^{S}_{i}$ is chosen because it does not commute with the instant mixer Hamiltonian $H_{M}^{S}(t)$ in {\it S} but commutes with the target Hamiltonian $H_{P}^{S}$ in {\it S}. 
For the most efficient ancillary operator in {\it A} subspace, we need to choose $\sum_{ij}X_{i}^{S} \otimes X_{j}^{A}$ or $\sum_{ij}X_{i}^{S} \otimes Z_{j}^{A}$ with the most pronounced landscape instead of the flattened one $\sum_{ij}X_{i}^{S}\otimes Y_{j}^{A}$, 
which requires a stronger kick angle to reach the same speedup. 

In practice, it is hard to determine precisely what the optimal kick angle value $\theta:=\theta_{opt}$ is to avoid excess excitations, and to have a reliable simulation at the end but to run experiments at different system sizes.
In Fig. 1, our numerical demonstration shows the periodic kick Hamiltonian $\theta(t=0^{+}, N_{K} = 43)\sum_{ij}X_{i}^{S} \otimes Z_{j}^{A}$
with $\theta_opt = 0.004$
spanning the duration $t \lesssim \tau$. The reduced system energy is calculated after tracing out ancillary DOF with the kicks. The true ground state energy is calculated with imaginary time evolution in {\it S-only} subspace.

In subplots~\ref{FIG:2}(a) and~\ref{FIG:2}(b) for the slower annealing $\Delta_{min} \tau \approx \theta_{XX}^{0}\tau = 0.5$, the system dynamics reaches the same reduced system energy faster.  As a byproduct, the speedup lifts the reduced system energy from the true ground-state energy floor at the end of the simulation.  
For the Floquet quantum kicks with larger $\theta$ (not shown), excited quantum states can lead to very wrong answers
at the end of the quantum simulation. For a much weaker kick angle $\theta \ll 0.004$ (not shown),
we cannot achieve any benefits from the kicks as opposed to the case without the kicks.

In subplots~\ref{FIG:2}(c) and~\ref{FIG:2}(d) with the faster annealing $\theta_{XX}^{0}\tau = 0.25$ and high-frequency kicks, the excess excitations are suppressed and lead to the right drift to the true ground state system energy~\cite{SM1.1}.
The time to the steady state for the reduced system energy is much shorter (one hundred percent reduction) signaling the acceleration to
the target ground state system energy.
The drawback from excess excitations is completely negated with the high-frequency ($\theta_{Z}^{0}\Delta t_{K} \ll 1$ and $\theta_{XX}^{0}\Delta t_{K} \ll 1$ ) quantum kicks. 

In subplots~\ref{FIG:2}(e)-\ref{FIG:2}(h) illustrated is the reduced system dynamics from the system Hamiltonian $H^{S}(t)$ and the continuous constant kick Hamiltonian $\theta(t)\sum_{ij}X^{S}_{i}Z^{A}_{j}$ spanning the whole simulation time $T$, the kicks always drive the system energy closer to the true system energy even at faster annealing. However, the speedup is still effective (with less error and faster convergence to final results than the case without kicks)
at a slower or faster annealingtime $\tau$ (as shown in subplots~\ref{FIG:2}(e-f) versus~\ref{FIG:2}(g-h)).
It is encouraging for the NN-TFIMs that the optimal kick angle is not sensitive to the system size for the quantum kick protocol.

{\it Infinite long-ranged transverse field Ising model (ILR-TFIM) in a closed chain.---}
To study the effect with a similar gap size near QPT as the NN-TFIM, we rescale the Ising interaction $\theta^{0}_{XX}$
to the lower value of $\theta_{XX}^{0}\rightarrow\theta_{XX}^{0}/(0.5N_S)$ while keeping the same $\theta_{Z}^{0}$, in which the scaling factor
$0.5N_S$ is given by the ratio of the connectivity between ILR-TFIM:$N_S(N_S-1)/2$ and NN-TFM: $N_S-1$.
The robustness of the quantum kick protocol prevails with the ILR-TFIM system  
Hamiltonian $ H^{S}_{P} = - \sum\limits_{i < j }\theta^{0}_{XX}X^{S}_{i}X^{S}_{j}$~\cite{SM4}.

{\it Hydrogen molecule model.---}
Due to the fermionic parity symmetry inherent to the low-energy valence electrons in a hydrogen molecule (${\rm{H_2}}$), the $\rm H_{2}$ quantum state is encoded by two spins after the Bravyi-Kitaev transformation~\cite{Bravi,Bravi2} and tapering~\cite{omalley2016,majumder2023}. The relevant low-energy subspace is spanned by $|01\rangle$ and $|10\rangle$ basis states. The system Hamiltonian is given by the linear combination of the Pauli terms $I^{S}_{0}I^{S}_{1}$, $Z^{S}_{0}Z^{S}_{1}$, $I^{S}_{0}Z^{S}_{1}$, $Z^{S}_{0}I^{S}_{1}$, and $X^{S}_{0}X^{S}_{1}+Y^{S}_{0}Y^{S}_{1}$ with given coupling interaction for different bond lengths shown in $\rm H_{2}$ data~\cite{SM3}. 
${\rm{H_2}}$ system Hamiltonian commutes with $\sum_{i}Z^{S}_{i}$ due to the parity conservation. If we initiate the quantum state at $|01\rangle$ or $|10\rangle$, the ${\rm{H_2}}$ Hamiltonian will not change parity, the eigenvalue of $\sum_{i}Z^{S}_{i}$.
By our proposed protocol, we need a mixer Hamiltonian not commuting with H$_2$ system Hamiltonian that can facilitate the superposition between states with different parities initially and does not lead us astray to the excited states.
Therefore, we need a one-local system mixer Hamiltonian that is not $\theta_{M}(t)\sum_{i}Z^{S}_{i}$ but the mixer $\theta_{M}(t)\sum_{i}X_{i}^{S} = \theta_{X}^{0}e^{-t/\tau}\sum_{i}X_{i}^{S}$ and carry out
the quantum simulation with annealing. For the choice of an efficient quantum kick, we select the instantaneous kick Hamiltonian commuting
with the relevant H$_2$ system Hamiltonian $X^{S}_{0}X^{S}_{1}+Y^{S}_{0}Y^{S}_{1}$, we can still speed up the quantum simulation as TFIMs and find the exact ground state energy for ${\rm H_{2}}$ with the protocol. 
As one observes~\cite{SM4}, this is not the most efficient kick but still reaches the simulation goal with a larger optimal kick angle $\theta$.
The efficient quantum kick unitary $U^{SA}(\hat {\Theta}^{A},\hat{S})$ for the $\rm H_2$ molecule model is chosen  
with the ancillary operator ${\hat {\Theta}}^{A} = \theta \sum_{l} Z_{l}^{A}$ and system operator $\hat{S} =\sum_{i}\frac{1}{\sqrt 2}(X_{i}^{S}+Y_{i}^{S})$ respectively.

\begin{figure}[h]
    \centering
    \includegraphics[scale=0.55]{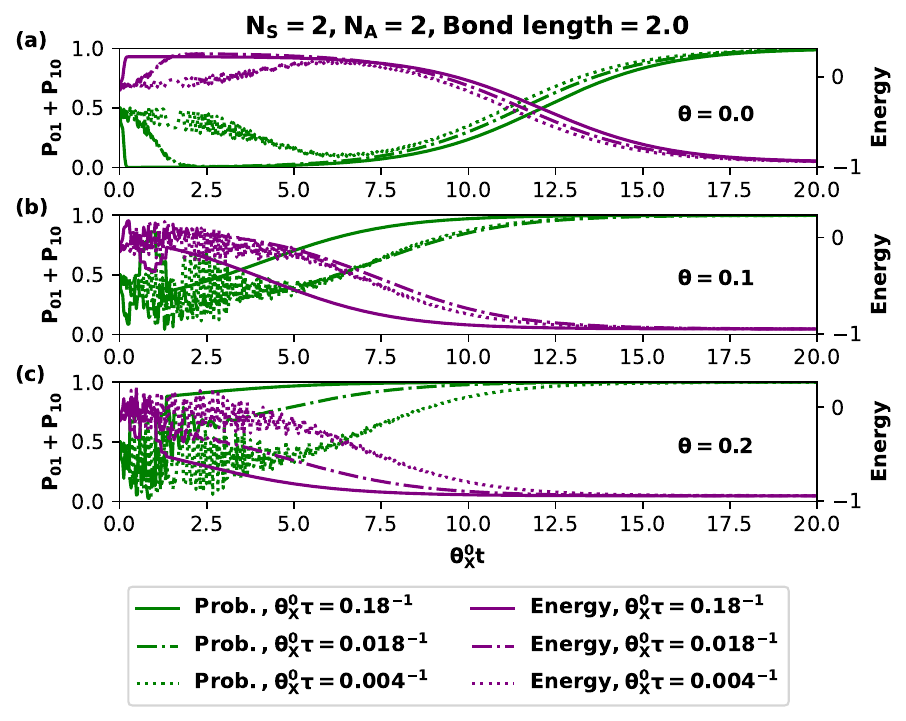}
    \caption{Time dependency of probability in odd parity subspace $|01\rangle$ and $|10\rangle$ (in green) and H$_2$ system binding energy (in purple) with $N_{S} = 2$, $N_{A} = 2$, bond length $= 2.0 a.u.$, and $\theta_{X}^{0} = 1.5$. The annealing rates $\theta_{X}^{0}\tau$ are shown in the attached legend. Different line types present cases with different annealing rates.
    The kick angle $\theta = 0$ labels the cases without the kicks
    in the top subplot (a). The tick values at the left vertical axis mark the values for the probability measurement cases (green), and those marked at the right vertical axis are for the energy measurement (purple). }
    \label{fig:FIG5}
\end{figure}

\begin{figure}[h]
    \centering
    \includegraphics[scale=0.55]{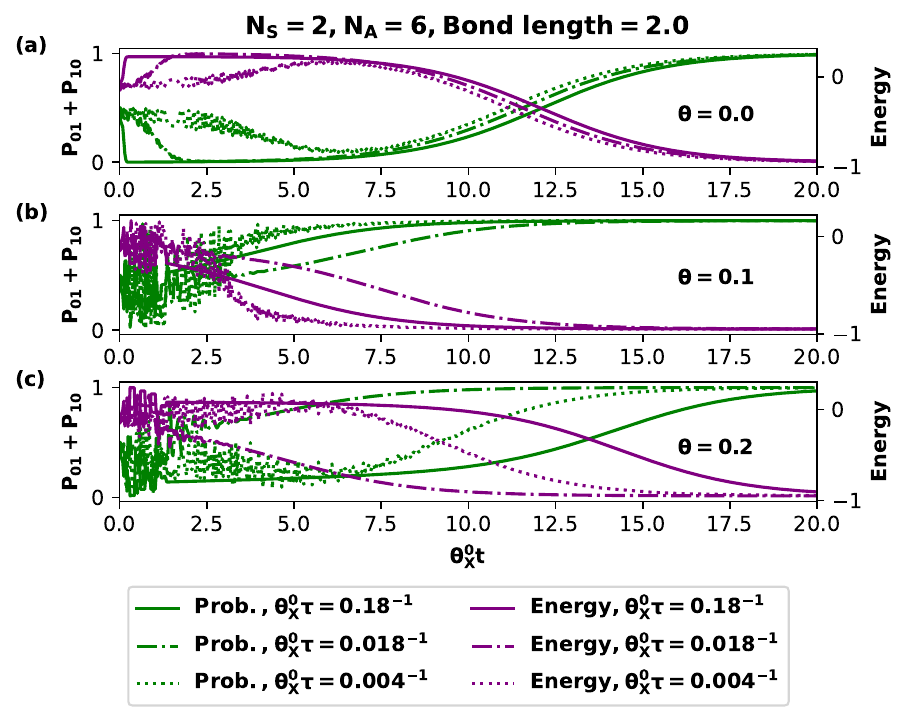}
    \caption{Time dependency of probability in odd parity subspace $|01\rangle$ and $|10\rangle$ and H$_2$ system binding energy with $N_{S} = 2$, $N_{A} = 6$, bond length $= 2.0$, and $\theta_{X}^{0} = 1.5$. The annealing rates $\theta_{X}^{0}\tau$ are shown in the attached legend. The cases without the kicks 
    are labeled by the kick angle $\theta = 0$ in the top subplot.}
    \label{fig:FIG6}
\end{figure}

Following the schedule Eq.~(\ref{EQ:2}) with the problem Hamiltonian $H^{S}_{P}$ replaced by the H$_2$ Hamiltonian, we illustrate our numerical speedup results in Fig.~\ref{fig:FIG5} and Fig.~\ref{fig:FIG6} for distinct 
bond length and interaction in atomic units~\cite{SM3}.
The full quantum state is initialized at the product state $|1_{x}\rangle^{\otimes 2}\bigotimes |1_{x}\rangle^{\otimes N_A}$, in which
$|1_{x}\rangle^{\otimes 2}$ is the ground state for the mixer Hamiltonian $\theta_{X}^{0}\sum_{i}X_{i}^{S}$.
It is noted that the sampled kick angles $\theta$ shown are not near optimal for any annealing rate. 

In Fig.~\ref{fig:FIG5} ($N_A = 2$, bond length $= 2.0~a.u$) and Fig.~\ref{fig:FIG6} ($N_A = 6$, bond length $= 2.0~a.u$), all purple energy curves collapse to the right binding energies at the long time limit. However, not all kick angles (fast or slow annealing rates) have the advantage over the case without the kick ($\theta = 0$), especially the case $\theta_{X}^{0}\tau=0.18^{-1}$ with fast annealing in Fig.~\ref{fig:FIG6}.
For the cases with hydrogen bond length $0.7~a.u.$~\cite{SM5},
we also observe the convergence to the true binding energy for any annealing rate. Still, with the clear advantages of the kicks, we show the kick angle dependence is not monotonic concerning the annealing rates due to the over-rotation of the same kick angles for excitations with more ancillary qubits.
We also observe the much shortened time $\theta_{X}^{0}t$ to the true system binding energy. 

In summary, we have developed a quantum speedup protocol using a digital-analog computing framework for quantum simulations bypassing the gap bottleneck of quantum simulation through adiabatic annealing. The protocol can be broadly applied for any quantum applications after Hamiltonian encoding including quantum sensing, quantum optimization~\cite{optimization}, quantum machine learning~\cite{PhysRevA.106.022601}, and quantum state preparation~\cite{Schuld}.

Our method demonstrates one hundred percent speedup (with respect to cases without kicks) in considered NN-TFIM, ILR-TFIM, and ${\rm H_{2}}$ models by applying quantum annealing algorithms with specific quantum kicks, particularly useful for simulations in noisy quantum hardware with short coherence time. Our design work also builds upon existing quantum algorithms and could be easily adapted to current or next-generation programmable digital-analog quantum devices in the lab. The quantum speedup primitive is novel from the standpoint of the modular algorithm design principle, without changing the core algorithms in the system space.

For an end-to-end quantum application chained through multiple Hamiltonian-encoded subroutines that function differently in an application such as state preparation or quantum information processing, the full algorithm comprises alternating annealing or ramping for the mixer and interested system Hamiltonians intersperse with quantum kicks for each subroutine layer. 
In each subroutine layer, we can reuse ancillary qubits, and reset the ancillary and system states at each functional layer by feedforward and measurement protocol over the ancillary subspace since the quantum information in the system subspace is preserved.
The ancillary resource remains constant for the increased complexity of the chained quantum subroutines.
Due to the environmental noise effects, with more intended effort, the primitive can potentially be rendered fault-tolerant by various schemes with further studies.

{\it Acknowledgments.--}
C.-C. Joseph Wang and Travis Humble acknowledge the support by the DOE Office of Science, Office of ASCR, under FWP No. ERKJ354 to initiate the project. The continuation of the project was supported by the U.S. Department of Energy, Office of Science, National Quantum Information Science
Research Centers, Quantum Science Center.

\section{Supplementary Material}
\renewcommand\theequation{S\arabic{equation}} 
\renewcommand\thefigure{S\arabic{figure}}
\section{I. Hydrogen molecule data}
\begin{table}[h]
\begin{center}
\caption{{$\rm{H_2}$~{\rm Data}.}Interaction coefficients for the Pauli Hamiltonian describing the H$_2$ molecule at two bond lengths.  The four-orbital (one-body and two-body) interactions were produced by the PySCF package~[52].  These were then mapped via the Brayvi-Kitaev transformation and tapering to a two-qubit Pauli Hamiltonian.}
\label{TABLE:1}
\begin{tabular}{||c | c | c ||} 
 \hline
    & Atomic units & Atomic units \\ \hline
 Bond length & 0.7       & 2.0       \\ \hline
 Exact system binding energy   & -1.136189  & -0.948641 \\ \hline
$I_{0}^{S}I_{1}^{S}$ coupling      & 0.304795   & -0.272905 \\ \hline
$I_{0}^{S}Z_{1}^{S}$ coupling      & 0.355426   & 0.134559  \\ \hline
$Z_{0}^{S}I_{1}^{S}$ coupling      & -0.485486  & 0.0133026 \\ \hline
$Z_{0}^{S}Z_{1}^{S}$ coupling      & 0.581232   & 0.389632  \\ \hline
$X_{0}^{S}X_{1}^{S}$ coupling      & 0.089500   & 0.129569  \\ \hline
$Y_{0}^{S}Y_{1}^{S}$ coupling      & 0.089500   & 0.129569  \\ [1ex] 
\hline
\end{tabular}
\end{center}
\end{table}

\section{II. Time-dependent Hamiltonian in the transverse field rotating frame}
In general, understanding quantum many-body dynamics with time-dependent Hamiltonian with noncommuting interactions is a difficult task. A powerful analytical approach is averaged Hamiltonian theory where the time-dependent Hamiltonian is replaced by an effective 
time-independent Hamiltonian. The effective Hamiltonian is subject to Magnus expansion term by term where each term still preserves the hermicity while being truncated at any order. With the strongest interaction as the rotating reference frame justifying the convergence of
truncated Magnus expansion, the prediction of the theory has been found useful for the description of the quantum many-body dynamics even to the lowest orders.
For a more detailed introduction, see Ref. [36] by Andreas Brinkmans at https://doi.org/10.1002/cmr.a.21414.
For a detailed review and applications, see reference [37] by S. Blanes et al..

Start with the system plus  ancillary Hamiltonian $H(t)$ in the Schrodinger picture
\begin{equation} 
H(t) = H_0(t) + H^{SA}(t) + H_{XX}, \ \  H_0(t) =\theta_M(t)\sum_{i}Z_{i}^{S}, \ \ H^{SA}(t) =  \theta(t)\sum_{ij}X_{i}^{S}Z_{j}^{A}, H_{XX} = \sum_{i<j}\theta_{XX}^{0} X_{i}X_{j}, 
\label{EQ:4}
\end{equation}
in which the dynamical kick angle is given by $\theta(t):= \theta \sum_{K=0}^{N_K-1}\delta(t-K\Delta t_K)$
with the kick time interval $\Delta t_{K}$ and $N_K$ is the total number of kicks after the evolution time $t$.
Then the Hamiltonian in the rotating frame picture is given by
\begin{align} H^{SA}(t) & \rightarrow \theta_M(t) \sum_{ij} e^{i \Theta_M(t) \sum_i Z_i^S} X_i^S Z_j^A  e^{-i \Theta_M(t) \sum_i Z_i^S} \nonumber\\
& = \theta_M(t) \sum_{ij}Z_j^A  e^{i \Theta_M(t) Z_i^S} X_i^S  e^{-i \Theta_M(t) Z_i^S} \end{align}
where $\Theta_M(t) = \int_0^t dt \theta_M(t)=\theta_{M}(0)\tau(1-e^{-t/\tau})$ and the second line simplifies by cancelling commuting terms.  After simplification, the terms $e^{i \Theta_M(t) Z_i^S} X_i^S Z_j^A  e^{-i \Theta_M(t) Z_i^S}$ are 
\begin{align} e^{i \Theta_M(t) Z_i^S} X_i^S  e^{-i \Theta_M(t) Z_i^S} & = \left[\cos(\Theta_M(t)) + i\sin(\Theta_M(t))Z_i\right]X_i^S  \left[\cos(\Theta_M(t)) - i\sin(\Theta_M(t))Z_i\right] \nonumber \\
& = \left[\cos^2(\Theta_M(t)) - \sin^2(\Theta_M(t))\right] X_i + 
i\cos(\Theta_M(t))\sin(\Theta_M(t))\left[ Z_i^S X_i^S - X_i^SZ_i^S\right] \nonumber\\
& = \cos(2\Theta_M(t)) X_i^S - \sin(2\Theta_M(t))Y_i^S 
\label{EQ:5}
\end{align}
Therefore, the expression for $H^{SA}(t)$ in the rotating frame is given by
\begin{equation} H^{SA}(t) = \theta_M(t) \sum_j Z_j^A\sum_{i} \cos(2\Theta_M(t)) X_i^S - \sin(2\Theta_M(t))Y_i^S. 
\label{EQ:6}
\end{equation}
In the rotating frame,  the static $H_{XX}$ becomes time-dependent given by the expression
\begin{equation} H_{XX}\rightarrow  H_{XX}(t) = \theta_{XX}^{0} \sum_{i < j} (\cos(2\Theta_M(t)) X_i - \sin(2\Theta_M(t)) Y_i) (\cos(2\Theta_M(t)) X_j - \sin(2\Theta_M(t)) Y_j). 
\label{EQ:7}
\end{equation}

\section{III. Time-averaged Hamiltonian}
Let us derive the leading time-averaged Hamiltonian for $H_{XX}(t)$ and $H^{SA}(t)$.
By direct substitution, the time-averaged Hamiltonian for $H_{XX}(t)$ is given by
\begin{equation}
\overline{H}_{XX}(T) \approx \int_{0}^{T}dt\frac{H_{XX}(t)}{T}
= -\theta_{XX}^{0}/T\left[\int_{0}^{T}dt\cos^2{2\Theta_{M}(t)} \sum_{i < j}X_{i}^{S}X_{j}^{S} + 
\int_{0}^{T}dt\sin^2{2\Theta_{M}(t)} \sum_{i < j}Y_{i}^{S}Y_{j}^{S} \right],
\label{EQ:8}
\end{equation}
in which the rapid oscillating $XY$ and $YX$ terms can be neglected.

For the Floquet kick Hamiltonian, 
\begin{equation}
H^{SA}(t) =  \theta(t)\left[\cos{2\Theta_{M}(t)} \sum_{ij}X_{i}^{S}Z_{j}^{A} - \sin{2\Theta_{M}(t)} \sum_{ij}Y_{i}^{S}Z_{j}^{A}\right].
\label{EQ:9}
\end{equation}
The time-averaged Hamiltonian is given by
\begin{equation}
\overline{H}^{SA}(T)=\int_{0}^{T}dt\frac{H^{SA}(t)}{T}
= 1/T\left[\int_{0}^{T}dt\theta(t)\cos{2\Theta_{M}(t)} \sum_{ij}X_{i}^{S}Z_{j}^{A} -\int_{0}^{T}dt\theta(t)\sin{2\Theta_{M}(t)} \sum_{ij}Y_{i}^{S}Z_{j}^{A}\right].
\label{EQ:10}
\end{equation}
We can further estimate the above sine integral
as
\begin{equation}
\int_{0}^{T}
dt\theta(t)\sin{2\Theta_{M}(t)} = \theta N_{K}
\sin\left[2\Theta_{M}(K\Delta t_{K})\right]
= \theta N_{K} \sin\left[2\theta_{M}(0)\tau(1-e^{-K\Delta t_{K}/\tau})\right]  
\nonumber
\end{equation}
\begin{equation}
\lesssim \theta N_{K}\sin(2\theta_{M}(0)\tau),
\label{EQ:11}
\end{equation}
in which $N_{K}$ is the accumulated number of kicks after time $T$.
This result works strictly for fast annealing$\tau \ll K\Delta t_{K}$
with the kick time interval $\Delta t_{K}$. However, it can generally be the upper bound for any annealingrates.
Similarly, the cosine integral can be estimated the same way.

\section{IV. State evolution from density matrix expansion}
Let us evaluate the state evolution by the full density matrix for our Floquet kick protocol in which  
$\overline{H}_{XX}(T)$ for the state evolution is negligible. The approximate full density matrix $\rho(T) \approx \sum_{n=0}^{2}\rho^{(i)}(T)$ 
from the truncated density matrix $\exp{(+iT\overline{H}^{SA})}\rho(0)\exp{(-iT\overline{H}^{SA})}$ under the effective unitary evolution $\exp{(-iT\overline{H}^{SA})}$ from the Floquet kicks are expanded to leading orders as follows. \\
For $n =0$, $\rho^{0}(T)$ is given by the density matrix from the initial state
\begin{equation}
\rho^{(0)}(T) = \rho(0).
\label{EQ:12}
\end{equation}
For $n=1$, the full-density matrix is given by
\begin{equation}
\rho^{(1)}(T) = -iN_{K}\theta \sin(2\theta_M(0)\tau)[\sum_{ij}Y_{i}^{S}Z_{j}^A, \rho(0)]
+i N_{K}\theta \cos{(2\theta_M(0)\tau)}[\sum_{ij}X_{i}^{S}Z_{j}^{A}, \rho(0)],
\label{EQ:13}
\end{equation}
in which the explicit $T$ dependence is absorbed in $N_{K}$. \\
For $n = 2$, the full density matrix is
given by
\begin{equation}
\rho^{(2)}(T) = -\frac{N_{K}^2 \theta^2 \cos^2(2\theta_M(0)\tau)}{2} \left[\sum_{lm} X_{l}^{S}Z_{m}^{A},\left[ \sum_{ij} X_{i}^{S}Z_{j}^{A}, \rho(0)\right]\right]  
\label{EQ:14}
\end{equation}

\begin{equation}
-\frac{N_{K}^2\theta^{2}\sin^{2}({2\theta_M(0)\tau})}{2} \left[\sum_{lm} Y_{l}^{S}Z_{m}^{A},\left[ \sum_{ij} Y_{i}^{S}Z_{j}^{A}, \rho(0)\right]\right]  
\label{EQ:15}
\end{equation}

\begin{equation}
+ \frac{N_{K}^2\theta^2 \sin(2\theta_M(0)\tau)\cos(2\theta_M(0)\tau)}{2}\left[\sum_{lm} X_{l}^{S}Z_{m}^{A},\left[ \sum_{ij} Y_{i}^{S}Z_{j}^{A}, \rho(0)\right]\right]
\label{EQ:16}
\end{equation}

\begin{equation}
+ \frac{N_{K}^2\theta^2 \sin(2\theta_M(0)\tau)\cos(2\theta_M(0)\tau)}{2}\left[\sum_{lm} Y_{l}^{S}Z_{m}^{A},\left[ \sum_{ij} X_{i}^{S}Z_{j}^{A}, \rho(0)\right]\right] 
\label{EQ:17}
\end{equation}

Let us write down the relevant terms in the target Hamiltonian $H_{XX}^{I}$ as
\begin{equation}
\sum_{l < m}X_{l}^{S}X_{m}^{S} = \sum_{l < m}\left(|0_{l}\rangle\langle 1_{l}|+|1_{l}\rangle\langle 0_{l}|\right)\left(|0_{m}\rangle\langle 1_{m}|+|1_{m}\rangle\langle 0_{m}|\right)\otimes I^{\notin l , m}.
\label{EQ:18}
\end{equation}

\begin{equation}
\sum_{l < m} Y_{l}^{S}Y_{m}^{S} = -\sum_{l < m}\left(|0_{l}\rangle\langle 1_{l}|-|1_{l}\rangle\langle 0_{l}|\right)\left(|0_{m}\rangle\langle 1_{m}|-|1_{m}\rangle\langle 0_{m}|\right)\otimes I^{\notin l , m}.
\label{EQ:19}
\end{equation}
 For details of the involved commutation relations for explicit calculation in the density matrix expansion, refer to the Sec. VI in the Supplementary Material.

Now, we are ready to evaluate the error bound for the target system energy $Tr_{S}\left(\rho_{S}(T)\overline {H}_{XX}(T)\right) - E_{S}^{T}:=\Delta E_{S} < \epsilon E_{S}^{T}$, 
in which the error $\epsilon$ is positive and the target true system energy $E_{S}^{T} < 0$ with respect to its corresponding system ground state $\rho_{Target}=|GND\rangle \langle GND|$ is defined by
\begin{equation}
Tr_{S}\left(\rho_{Target}\sum_{l < m}-\theta_{XX}^{0}X_{l}^{S}X_{m}^{S}\right):= E_{S}^{T}.
\label{EQ:20}
\end{equation}

In the following, we address the mathematical reason why the quantum kicks move the system energy toward the true system ground state for the cases with quantum speedup over the cases without the kicks.
We estimate the system energy $Tr_{S}\left(\rho_{S}(T)\overline {H}_{XX}\right)$ by following quantum state evolution mainly from the transverse field and quantum kicks for our protocol since 
the target system Hamiltonian $H_{XX}$ is not the leading-order driver for the state evolution before quantum phase transition (QPT) and the state change driven by $H_{XX}$ is perturbative for the later time dynamics (this is typical the cases with quantum speedup in our numerical results as long as the kick angle is not too large).
By Eqs. (\ref{EQ:12}), (\ref{EQ:13}), (\ref{EQ:18}), and (\ref{EQ:19}), we find no contribution to the system energy from leading quantum state dynamics  $E_{S}^{(0)}:= 
Tr_{S}\left(\rho^{(0)}_{S}(T)\overline {H}_{XX}(T)\right) =
0$, and  $E_{S}^{(1)}:=Tr_{S}\left(\rho^{(1)}_{S}(T)\overline {H}_{XX}(T)\right)
= 0$ because {\it only} off-diagonal maps are involved in the operators $\rho^{(0)}_{S}(T)\overline {H}_{XX}$ and $\rho^{(1)}_{S}(T)\overline {H}_{XX}$(T)
. 
However, we can identify diagonal maps contributing to $E_S^{(2)}:=Tr_{S}\left(\rho^{(2)}_{S}(T)\overline {H}_{XX}(T)\right)$. 
We find {\it only} Eqs.~(\ref{EQ:14}) and~(\ref{EQ:15}) contribute to the system energy after tracing over system basis states
\begin{equation}
E_S^{(2)} = -\theta_{XX}^{0}\frac{N_A^2 N_{K}^2 \theta^2 }{2} \left|\sum_{i < j}\right| F(T)
\label{EQ:21}
\end{equation}
in which the connectivity of the interactions is given by $\left|\sum_{i < j}\right|$,  and the $X^{S}X^{S}$, $Y^{S}Y^{S}$ terms (Eq.~(\ref{EQ:18}), Eq.~(\ref{EQ:19})) both contribute to the system energy.
The sign of the factor $F(\tau)$ determines the gain or loss in the system energy from
the chosen kicks in TFIM as given by the expression
\begin{equation}
F(T):=\sin^{2}(2\theta_M(0)\tau)\int_{0}^{T/\tau}dt'\cos^{2}2\Theta_{M}(t') + \cos^{2}(2\theta_M(0)\tau)\int_{0}^{T/\tau}dt'\sin^{2}2\Theta_{M}(t'),
\label{EQ:23}
\end{equation}
in which $\Theta_M(t')=\theta_{M}(0)\tau(1-\exp(-t'))$ from our annealing protocol with $\theta_{M}(0)\tau \gtrsim 1$ and the system energy analysis is valid for the simulation time $T$ is well greater than $\tau$ for our interest. We can see immediately the sign for the factor $F(T)$ is always positive. 
We find the kick will always cause the system energy flow $\Delta E_{S} = N_{A}^2 N_{K}^2 \theta^2 F(T) \propto E_{S}^{T}< 0$ synchronous with the sign of the ground state energy, lowering the system energy toward true ground state energy in the NN-TFIM cases (Fig.~1) and ILR-TFM cases (Fig. S3). Notice that the assumption that the contribution of state evolution from $H_{XX}$ should remain valid when $T \gg \tau$ especially when the kicks are optimal or sufficiently efficient by the numerical results shown in the letter.

Now we switch gears to understand the kick effects on the earlier system dynamics before the saturation of the system energy.
Now we evaluate the instant system's energy 
$Tr_{S}(\rho_{S}(T)\theta_{M}(t=T)\sum_{i}Z_{i}^{S})$
from the mixer Hamiltonian in the transverse field 
$\theta_{M}(t)\sum_{i} Z_{i}^{S}$ where $\rho_{S}(T)$ is from the trace of ancillary DOF
from the full density matrix $\rho(T)$. The energy is invariant in the rotating frame.
Only diagonal terms from $\rho_S(T)$ can contribute to the system energy with this Hamiltonian. This includes
$\rho^{(0)}(T)$ in Eq.~(\ref{EQ:12}) and $\rho^{(2)}(T)$ in Eqs.~(\ref{EQ:14}) and (\ref{EQ:15}).
The contribution from $\rho^{(0)}(T)$ is given by
\begin{equation}
E^{(0)}_{S}(T) = - N_{S}\theta_{M}(T).
\label{EQ:26}
\end{equation}
The crucial contribution from $\rho^{(2)}(T)$ is accordingly derived by substituting off-diagonal terms from Eqs. (\ref{EQ:35}) and (\ref{EQ:38}) to Eqs. (\ref{EQ:14}) and (\ref{EQ:15})
after tracing over ancillary DOF as
\begin{equation}
E^{(2)}_{S}(T) =  + N_{A}^2 N_{K}^2 \theta^2 N_{S} \theta_{M}(T).
\label{EQ:27}
\end{equation}
This positive contribution from below to the system energy $E_{S}(T) = E_{S}^{(0)}(T) + E_{S}^{(2)}(T)$ before QPT explains the speedup from our numerical results in
Fig. 1 and Fig. S3.
Formally, we can estimate the time $T^{*}$ to reach the ground state
energy $E_{S}(T^{*})-E_{S}^{T} < -\epsilon E_{S}^{T}$ with the error $\epsilon > 0$ from below in the presence of the Floquet kicks ($\theta \neq 0$) and the absence of the Floquet kicks($\theta = 0$).
Therefore, the optimal kick angle $\theta_{opt}$ with no error $\epsilon = 0$ is estimated asymptotically by $E_{S}(T) = E_{S}^{T}$ after the time $T$ can be derived
as
\begin{equation}
\theta_{opt} \approx \frac{1}{N_A N_{K}}\sqrt{1- E_{S}^{T}/E_{S}^{Mixer}(T)}.
\label{EQ:28}
\end{equation}
On the other hand, the ratio of the time to reach the target system energy with kicks $\theta \neq 0$ and without the kicks $\theta = 0$ defines
the speedup as
\begin{equation}
T^{*}(\epsilon ,\theta \neq 0)/T^{*}(\epsilon, \theta=0) = 1+ \frac{\ln\left[1-\theta^2N_A^2N_{K}^2\right]}{\ln{\left[- N_{S}\theta_{M}(0)/(1-\epsilon)E_{S}^{T}\right]}},
\label{EQ:29}
\end{equation}
in which $\epsilon > 0$ and $E_{S}^{T}\theta_{M}(0)< 0$ and $-N_S\theta_{M}(0)/E_{S}^{T}\gg 1$.
This agrees with our numerical results, this ratio is always less than one and
can be one-hundred percent speedup as our numerical results.
For the optimal angle $\theta_{opt}$, the speedup ratio can be estimated by replacing $\theta$ by $\theta_{opt}$. 

\section{V. Fidelity}
By careful evaluation, only the off-diagonal terms in Eq.~(\ref{EQ:13}) and Eq.~(\ref{EQ:15}) in the time-averaged density matrix expansion 
will contribute to the GHZ state fidelity $\mathscr{F}:=\langle GHZ|\rho_{S}(T)|GHZ\rangle$ due to the parity conservation for the $GHZ$ state $|GHZ\rangle = 2^{-1/2}(|+\rangle^{\otimes N_S}+|-\rangle^{\otimes N_S})$.
Therefore, we derive the increasing fidelity
from Eqs.~(\ref{EQ:33}) and (\ref{EQ:35}) respectively toward the $GHZ$ state fidelity by changing to the correct sign quadratically as
\begin{equation}
\mathscr{F}_{K} = 2^{-N_S+1}N_S N_A \theta N_{K} \sin(2\theta_{M}(0)\tau) + 2^{-N_S+1}N_S N_A^2 N_{K}^2 \theta^2 \sin^{2}(2\theta_{M}(0)\tau).
\label{EQ:30}
\end{equation}
Notice this confirms the fidelity to the ground state $|GHZ\rangle$ is improved by quantum kicks.

Let us next evaluate the fidelity distance to the target state without the kicks($\theta = 0$).
For the zeroth order in which $\rho_{S}^{(0)}=|1\rangle^{\otimes N_S}\langle 1|^{\otimes N_S}$, the fidelity is 
\begin{equation}
\mathscr{F}^{0} =\langle GHZ|\rho_{S}^{(0)}|GHZ\rangle = 2^{-N_S+1} \vee 0.0
\label{EQ:31}
\end{equation}
for even $\vee$ odd system qubits.

We {\it only} need to consider Eq.~(\ref{EQ:12}) in the approximate full-density expansion without the kicks. One can carry out the fidelity calculation and find out $XX$
and $YY$ terms in $\overline{H}_{XX}(T)$
do not contribute to the density matrix $\rho^{(0)}(T)$ and the fidelity without the kick is negligible  
\begin{equation}
\mathscr{F}_{XX} = \langle GHZ| Tr_{A} \left[ -iT\overline{H}_{XX}(T),\rho(0) \right] |GHZ\rangle \approx 0. 
\label{EQ:32}
\end{equation}
This is consistent with dropping the negligible $H_{XX}$ term for the quantum state evolution before the quantum phase transition in TFIM.  

\section{VI. Identities}
Let us summarize the following useful identities
\begin{align}
&\left[\sum_{ij}Y_{i}^{S}Z_{j}^{A}, \rho(0)\right] =
\left[-i\sum_{i}(|0_{i}\rangle\langle 1_{i}|-|1_{i}\rangle\langle 0_{i}|)\bigotimes \sum_{j}|0_{j}\rangle\langle 0_{j}|-|1_{j}\rangle\langle 1_{j}|, |1\rangle^{\otimes N_{S}}\langle 1|^{\otimes N_{S}} \bigotimes |0\rangle^{\otimes N_{A}}\langle 0|^{\otimes N_{A}}\right] \ \ \ \ \ \ \ \ \nonumber \\
& = -i N_{A} \sum_{i}(|1_{i}\rangle\langle 0_{i}|+|0_{i}\rangle\langle 1_{i}|)\otimes|1_{\neq i}\rangle^{\otimes N_{S}-1}\langle 1_{\neq i}|^{\otimes N_{S}-1} \bigotimes |0\rangle^{\otimes N_{A}}\langle 0|^{\otimes N_{A}}.
\label{EQ:33}
\end{align}

\begin{align}
& \left[\sum_{lm}Y_{l}^{S}Z_{m}^{A}, \left[\sum_{ij}Y_{i}^{S}Z_{j}^{A},\rho(0)\right]\right] = -2N_{A}^{2}\sum_{i}(|0_{i}\rangle\langle 0_{i}|-|1_{i}\rangle\langle 1_{i}|)
\otimes|1_{\neq i}\rangle^{\otimes N_{S}-1}\langle 1_{\neq i}|^{\otimes N_{S}-1} \bigotimes |0\rangle^{\otimes N_{A}}\langle 0|^{N_{A}}, \ \ \ \ \ \ \ \ \  \nonumber \\
& - N_{A}^{2} \sum_{l}\sum_{i \neq l}(|1_{i}\rangle\langle 0_{i}|+|0_{i}\rangle\langle 1_{i}|)\otimes (|0_{l}\rangle\langle 1_{l}|+|1_{l}\rangle\langle 0_{l}|)\otimes |1_{\neq i,l}\rangle^{\otimes N_S-2}\langle 1_{\neq i,l}|^{\otimes N_S -2} \bigotimes
|0\rangle^{\otimes N_A}\langle 0|^{N_A}. 
\label{EQ:35}
\end{align}

\begin{align}
& \left[\sum_{ij}X_{i}^{S}Z_{j}^{A}, \rho(0) \right] & = - N_{A} \sum_{i}  \left[
(|1_{i}\rangle\langle 0_{i}|-|0_{i}\rangle\langle 1_{i}|)\otimes |1_{\neq i}\rangle^{\otimes N_S-1} \langle 1_{\neq i}|^{\otimes N_S-1}
\right] \bigotimes |0\rangle^{\otimes N_A}\langle 0|^{\otimes N_A}. \ \ \ \ \ \ \ \ \
\ \ \ \ \ \ \ \ \ \ \ \ \ \ 
\label{EQ:37}
\end{align}

\begin{align}
& \left[\sum_{lm} X_{l}^{S}Z_{m}^{A},\left[ \sum_{ij} X_{i}^{S}Z_{j}^{A}, \rho(0)\right]\right]  =
 - 2N_{A}^2\left[\sum_{l}(|0_l\rangle\langle 0_{l}|- |1_l\rangle\langle 1_{l}| )\otimes |1_{\neq l}\rangle^{\otimes N_S-1}\langle 1_{\neq l}|^{\otimes N_S-1}\right] \bigotimes |0\rangle^{\otimes N_{A}}\langle 0|^{\otimes N_{A}} \ \ \ \ \ \ \nonumber \\
& + N_{A}^2\left[\sum_{l}\sum_{i\neq l}(|1_i\rangle\langle 0_{i}|- |0_i\rangle\langle 1_{i}| )\otimes 
(|1_{l}\rangle\langle 0_{l}|-|0_{l}\rangle\langle 1_{l}|)
\otimes|1_{\neq i,l}\rangle^{\otimes N_S-2}\langle 1_{\neq i,l}|^{\otimes N_S-2}\right]
\bigotimes |0\rangle^{\otimes N_{A}}\langle 0|^{\otimes N_{A}}. 
\label{EQ:38}
\end{align}

\begin{align}
& \left[\sum_{lm} Y_{l}^{S}Z_{m}^{A},\left[ \sum_{ij} X_{i}^{S}Z_{j}^{A}, \rho(0)\right]\right] =
-i N_{A}^{2}\sum_{l}\sum_{i\neq l}(|1_{i}\rangle\langle 0_{i}|-|0_{i}\rangle \langle 1_{i}|)\otimes
(|1_{l}\rangle\langle 0_{l}|+|0_{l}\rangle\langle 1_{l}|)
\otimes |1_{\neq i,l}\rangle^{N_S-2}
\langle 1_{\neq i,l}|^{N_S-2} \ \ \ \nonumber \\   
& \bigotimes |0\rangle^{\otimes N_{A}}
\langle 0|^{\otimes N_A}.
\label{EQ:40}
\end{align}

\begin{align}
& \left[\sum_{ij} X_{i}^{S}Z_{j}^{A},\left[ \sum_{lm} Y_{l}^{S}Z_{m}^{A}, \rho(0)\right]\right] = -iN_{A}^{2}\sum_{l}\sum_{i\neq l}(|0_{l}\rangle\langle 1_{l}|+|1_{l}\rangle\langle 0_{l}|)
\otimes
(|0_{i}\rangle\langle 1_{i}|-|1_{i}\rangle\langle 0_{i}|)
\otimes |1_{\neq i,l}\rangle^{\otimes N_S-2}\langle 1_{\neq i,l}|^{\otimes N_S-2} \nonumber \\
& \bigotimes
|0\rangle^{\otimes N_A}
\langle 0|^{\otimes N_A}.
\label{EQ:41}
\end{align}

\begin{align}
& \left[ \sum_{l\neq m}Y_{l}^{S}X_{m}^{S},\rho(0)\right]
= i\sum_{l\neq m}(|1_{l}\rangle\langle 0_{l}|+|0_{l}\rangle\langle 1_{l}|)\otimes(|1_{m}\rangle\langle 0_{m}|+|0_{m}\rangle\langle 1_{m}|)
\otimes |1_{\neq l, m}\rangle^{\otimes N_{S}-2}\langle 1_{\neq l, m}|^{\otimes N_{S}-2}
\bigotimes  |0\rangle^{\otimes N_{A}}\langle 0|^{\otimes N_{A}}.
\label{EQ:42}
\end{align}

\begin{align}
& \left[\sum_{l \neq m}Y_{l}^{S}X_{m}^{S},\left[\sum_{i \neq j} Y_{i}^{S}X_{j}^{S},\rho(0)\right] \right]
 = \{ -4\sum_{l\neq m}(|0_{l}\rangle \langle 0_{l}|-|1_{l}\rangle \langle 1_{l}|)
\otimes 
(|0_{m}\rangle \langle 0_{m}|+|1_{m}\rangle \langle 1_{m}|)
\otimes 
|1_{\neq l, m}\rangle^{\otimes N_S-2}\langle 1_{\neq l, m}|^{\otimes N_S-2} \nonumber \\
& -2\sum_{l\neq m \neq j}(|0_{l}\rangle\langle 0_{l}|)\otimes
(|0_{m}\rangle \langle 1_{m}|+|1_{m}\rangle\langle 0_{m}|)\otimes
(|1_{j}\rangle\langle 0_{j}|+|0_{j}\rangle\langle 1_{j}|)\otimes |1_{\neq l,m,j}\rangle^{\otimes N_{S}-3}
\langle 1_{\neq l,m,j}|^{\otimes N_S-3} \nonumber \\
&
-\sum_{l\neq m \neq i \neq j}
[(|1_{l}\rangle \langle 0_{l}|\otimes |1_{m}\rangle\langle 0_{m}|)
\otimes
(|1_{i}\rangle\langle 0_{i}|+|0_{i}\rangle\langle 1_{i}|)
\otimes
(|1_{j}\rangle\langle 0_{j}|+|0_{j}\rangle\langle 1_{j}|) + 0\leftrightarrow 1]
\otimes |1_{\neq l,m,i,j,}\rangle^{\otimes N_{S}-4}
\langle 1_{\neq l,m,i,j}|^{\otimes N_S-4}\}  \nonumber \\
& \bigotimes |0\rangle^{\otimes N_A}\langle 0|^{\otimes N_A}.
\label{EQ:43}
\end{align}

\begin{align}
& \left[ \sum_{l\neq m}Y_{l}^{S}X_{m}^{S},\left[\sum_{ij} Y_{i}^{S}Z_{j}^{A}, \rho(0)\right]\right]=
\{-2N_{A}\sum_{l\neq m}(|0_{l}\rangle\langle 0_{l}|)\otimes(|1_m\rangle\langle 0_{m}|+|0_{m}\rangle\langle 1_{m}|)
\otimes |1_{\neq l, m}\rangle^{\otimes N_S-2}\langle 1_{\neq l,m}|^{\otimes N_S-2} \ \ \ \ \ \ \ \nonumber \\
& -N_{A}\sum_{i\neq l \neq m} (|1_{i}\rangle\langle 0_{i}|+|0_{i}\rangle \langle 1_{i}|)
\otimes
(|1_{l}\rangle\langle 0_{l}|\otimes|1_{m}\rangle\langle 0_{m}|+ 1 \leftrightarrow 0 )
\otimes
|1_{\neq i,l,m}\rangle^{\otimes N_S-3}\langle 1_{\neq i,l,m}|^{\otimes N_S-3} \}
\bigotimes 
|0\rangle^{\otimes N_A}\langle 0|^{\otimes N_A}.
\label{EQ:46}
\end{align}

\begin{align}
& \left[ \sum_{lm}Y_{l}^{S}Z_{m}^{A},
\left[\sum_{i \neq j} Y_{i}^{S}X_{j}^{S}, \rho(0)\right]\right]=\{
-4N_A\sum_{i \neq j} (|1_{i}\rangle\langle 0_{i}|+|0_{i}\rangle\langle 1_{i}|)
\otimes
(|0_{j}\rangle\langle 0_{j}|-|1_{j}\rangle\langle 1_{j}|)
\otimes
|1_{\neq i,j}\rangle^{\otimes N_S-2}\langle 1_{\neq i,j}|^{\otimes N_S-2}
\nonumber \\
& -2N_{A}\sum_{l\neq i \neq j}
(|0_{i}\rangle\langle 1_{i}|+|1_{i}\rangle \langle 0_{i}|)
\otimes
(|0_{j}\rangle\langle 1_{j}|+|1_{j}\rangle \langle 0_{j}|)
\otimes 
(|0_{l}\rangle \langle 1_{l}|+|1_{l}\rangle\langle 0_{l}|)
\otimes
|1_{\neq l,i,j}\rangle^{\otimes N_S-3}\langle 1_{\neq l,i,j}|^{\otimes N_S-3}\}
\bigotimes |0\rangle^{\otimes N_A}\langle 0|^{\otimes N_A}.
\label{EQ:47}
\end{align}

\begin{align}
& \left[ \sum_{l\neq m}Y_{l}^{S}X_{m}^{S},\left[\sum_{ij} X_{i}^{S}Z_{j}^{A}, \rho(0)\right]\right] 
= \{-2iN_{A}\sum_{l\neq m} (|0_{l}\rangle\langle 0_{l}|)
\otimes
 (|0_{m}\rangle\langle 1_{m}|-|1_{m}\rangle\langle 0_{m}|)
 \otimes
 |1_{\neq l,m}\rangle^{\otimes N_S-2}\langle 1_{\neq l,m}|^{\otimes N_S-2}
\nonumber \\
& -iN_{A}\sum_{i\neq l \neq m}(|1_{i}\rangle\langle 0_{i}|-|0_{i}\rangle\langle 1_{i}|)\otimes
(|1_{l}\rangle\langle 0_{l}|\otimes |1_{m}\rangle\langle 0_{m}|+ 1 \leftrightarrow 0)
\otimes |1_{\neq i,l,m}\rangle^{\otimes N_S-3}\langle 1_{\neq i,l,m}|^{\otimes N_S-3}\}
\bigotimes |0\rangle^{\otimes N_A}\langle 0|^{\otimes N_A}.
\label{EQ:49}
\end{align}

\begin{align}
&\left[ \sum_{lm}X_{l}^{S}Z_{m}^{A},
\left[\sum_{i \neq j} Y_{i}^{S}X_{j}^{S}, \rho(0)\right]\right] = 
-iN_{A}\sum_{i \neq j \neq l}(|1_{i}\rangle\langle 0_{i}|+|0_{i}\rangle\langle 1_{i}|)
\otimes
(|1_{j}\rangle\langle 0_{j}|+|0_{j}\rangle\langle 1_{j}|)
\otimes
(|1_{l}\rangle\langle 0_{l}|-|0_{l}\rangle\langle 1_{l}|) \ \ \ \ \ \ \ \ \nonumber \\
& \otimes 
|1_{\neq i,j,l}\rangle^{\otimes N_S-3}\langle 1_{\neq i,j,l}|^{\otimes N_S-3}
\bigotimes
|0\rangle^{\otimes N_A}\langle 0|^{\otimes N_A}.
\label{EQ:50}
\end{align}

\section{VII. Supplementary numerical results} 
\subsubsection*{Optimal kick angle}
Take NN-TFIMs in Fig.~1(c) and Fig.~1(d) for instance with the same number of kicks $N_{K}=43$ and ancillary qubits $N_A = 4$ but different number of system qubits $N_S$, the energy ratio is given by $E_{S}^{T}/E_{S}^{Mixer}(T\approx \tau = 0.25)\approx e(N_S-1)\theta_{XX}^{0}/N_S\theta_{Z}^{0}$.
Therefore, the optimal angle is insensitive to $N_S$ by the appearance of the $N_S$ both in the denominator and numerator of the above fraction and is given by the same  $\theta_{opt}\approx 0.004$, in excellent agreement with our numerical optimal angle $0.004$ by grid search. The same scenario applied to the ILR-TFIM. In
Figs.~\ref{fig:FIG3}(c) and~\ref{fig:FIG3}(d), we get the same approximate optimal angle $0.004$ since we
have chosen a much weaker $\theta_{XX}^{0}$ while maintaining similar $E_S^{T}/E_S^{Mixer}(T=0)$ for the long-ranged TFIM.
For the H$_2$ molecule cases in Figs.~2 and~3 with different $N_A=2$ and $N_A=6$, the energy ratios $E_{S}^{T}/E_{S}^{Mixer}(\tau)$ is approximately $(0.13/1.5)e\approx 0.236$ for the bond length $2.0~a.u.$.
We arrive at the estimated optimal angles as $\theta_{opt} \approx \frac{1}{2 N_K}(1-0.236)^{0.5} = 0.2185$ for $N_A=2$ and
$\theta_{opt} \approx \frac{1}{6 N_K}(1-0.236)^{0.5} = 0.073$ for $N_A=6$ (the same $N_K = 10$). This is close to the respective near-optimal angle $\theta=0.2$ in Fig.~2 and the near-optimal angle $\theta = 0.1$ in Fig.~3, as demonstrated by the faster saturation to the system ground state energy  (optimal speedup) from the purple curves at different annealing rates.

In Figs.~\ref{fig:FIG7} and~\ref{fig:FIG8}, the energy ratios $E_{S}^{T}/E_{S}^{Mixer}(T\approx \tau)$ equals approximately to $(0.09/1.5)e\approx 0.163$ for the bond length $0.7~a.u.$.
Accordingly, the estimated optimal angles as $\theta_{opt} \approx \frac{1}{20}(1-0.163)^{0.5} = 0.046 $ for $N_A = 2, N_K = 10$ and
$\theta_{opt} \approx \frac{1}{60}(1-0.163)^{0.5} = 0.015$ for $N_A = 6, N_K = 10$. The theory estimation does not agree well with
the expected numerical results due to the parametric assumption for the theory estimation.
The assumption that only the $H^{SA}(t)$ term and the mixer $H_{S}^{Mixer}(t)$ are the leading contributions
to the quantum state evolution with the negligible system Hamiltonian of interest, as manifested by the energy scale from terms other than $X^{S}X^{S}$ and $Y^{S}Y^{S}$ in H$_2$ data Table I. However, one can remedy this drawback by starting the annealing field $\theta_{X}^{0}$ at a much higher value, so that the theoretical estimation from Eq. (3) is reliable. 

Let us comment on how the optimal kick angle $\theta_{opt}$ scales with the system qubit number $N_S$ and the ancillary qubit count $N_A$.
From Eq. (3), the optimal angle is inversely proportional to $N_{A}$ and $N_K$ derived from the prefactor $(N_K N_A)^{-1}$.
For the system size dependence of the optimal kick angle, $\theta_{opt}$ is determined by the model-specific scaling from
$\sqrt{1-E^{T}_{S}(N_S)/E_S^{Mixer}(\tau, N_S)}$. For the NN-TFIM, $\theta_{opt}$ does not scale
with $N_S$ since  $E_{S}^{T}$ and $E_S^{Mixer}$ are all scale linearly with $N_S$.
For the ILR-TFIM, $\theta_{opt}$ scales as $\sqrt{N_S}$ because the true system ground state energy  $E_{S}^{T}$ scales
as $N_S(N_S-1)$ due to the infinite connection and the mixer system energy $E_S^{Mixer}$ scales normally as $N_S$
due to the one-local nature of the mixer Hamiltonian.

\section{VIII. Supplementary figures}

\begin{figure}[h]
    \centering
    \includegraphics[scale=0.65]{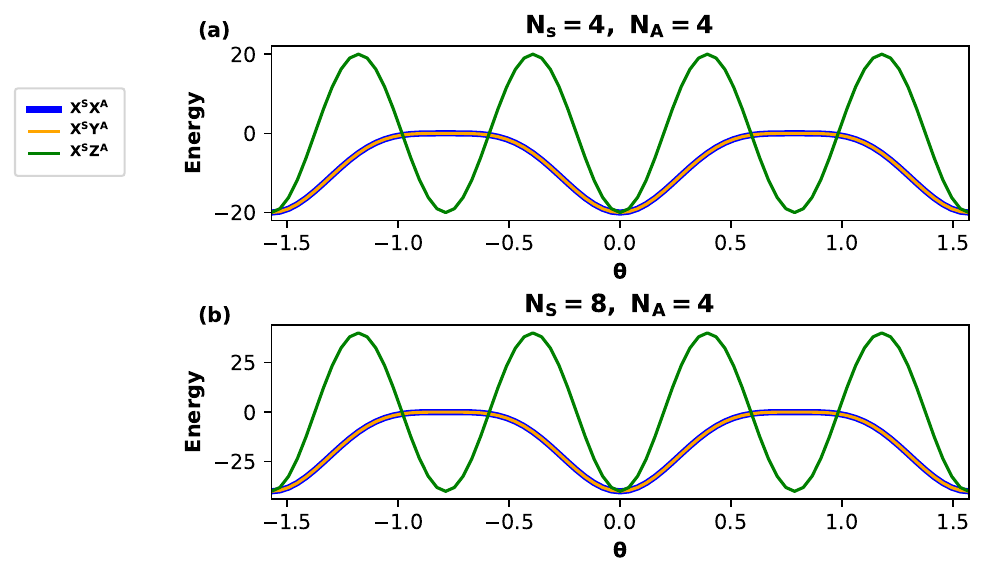}
    \caption{Reduced System energy (labeled by Energy) versus the kick angle $\bf \theta$ for a kick at $t=0$.
    Kick Hamiltonian types: $X^{S}X^{A}$(in green), $X^{S}Y^{A}$(in orange), and $X^{S}Z^{A}$(in thick blue). The initial system energy landscape is insensitive to the annealing rates and the nature of the models (NN-TFIM or ILR-TFIM). 
    (a) $N_{S} = 4, N_{A} = 4$, (b) $N_{S} = 8, N_{A} = 4$.}
    \label{FIG:1}
\end{figure}

As shown in Fig.~\ref{FIG:1} is the NN-TFIM system energy response to a quantum kick applied at $t=0$ versus kick angle $\theta$. The landscape is robust regardless of the annealing time scale $\tau$ because the quantum state is mostly dominated by the transverse field $H^{S}_{M}(t)$ ground state with a large gap. It is also not sensitive to the number of the system qubits $N_S$ but sensitive to the number of ancillary qubits $N_A$. The minimal required number for ancillary qubits $N_A$ with pronounced sensitivity is $4$ and larger $N_A$ than necessary does not provide more advantages from our studies in general.
The most efficient kick is of type $X^{S}Z^{A}$.
The less efficient kicks types $X^{S}X^{A}$, $X^{S}Y^{A}$ perform equally with overlapping landscapes.

\begin{figure}[h]
    \centering
    \includegraphics[scale=0.65]{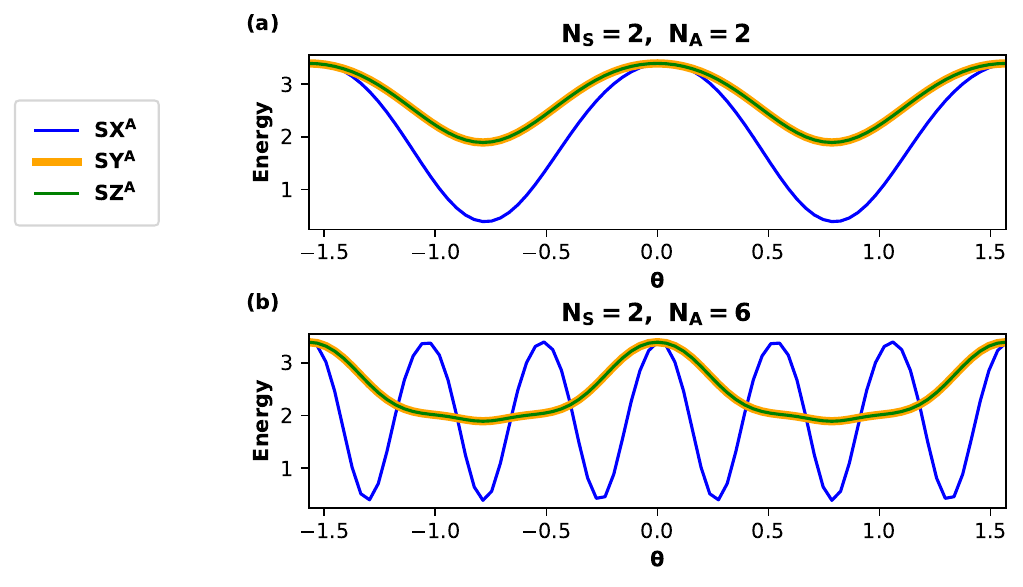}
    \caption{Hydrogen molecule system energy (Energy) versus kick angle $\bf \theta$ for a kick with angle $\theta$.
    Kick Hamiltonian types: $SX^{A}$(in blue), $SY^{A}$(in thick orange), and $SZ^{A}$(in green) is insensitive to different annealing rates $\theta_X^{0}\tau$.}
    \label{fig:FIG4}
\end{figure}

In Fig.~\ref{fig:FIG4}, the kick energy landscape for a hydrogen molecule versus the kick angle $\theta$ is robust at the time $t=0$ regardless of the annealing time scale $\tau$ because the quantum state is mostly dominated by the transverse field $H^{S}_{M}(t)$ ground state with a large gap. The landscape is insensitive to the number of the system qubits $N_S$ but is sensitive to the number of ancillary qubits $N_A$. The kick energy landscape has negative curvature near $\theta=0$ due
to the difference in the system Hamiltonian of $\rm {H_2}$. The most efficient kick is of type $SX^{A}$. The kicks $SY^{A}$ and $SZ^{A}$ are equal performers.

\begin{figure}[h]
    \centering
    \includegraphics[scale=0.75]{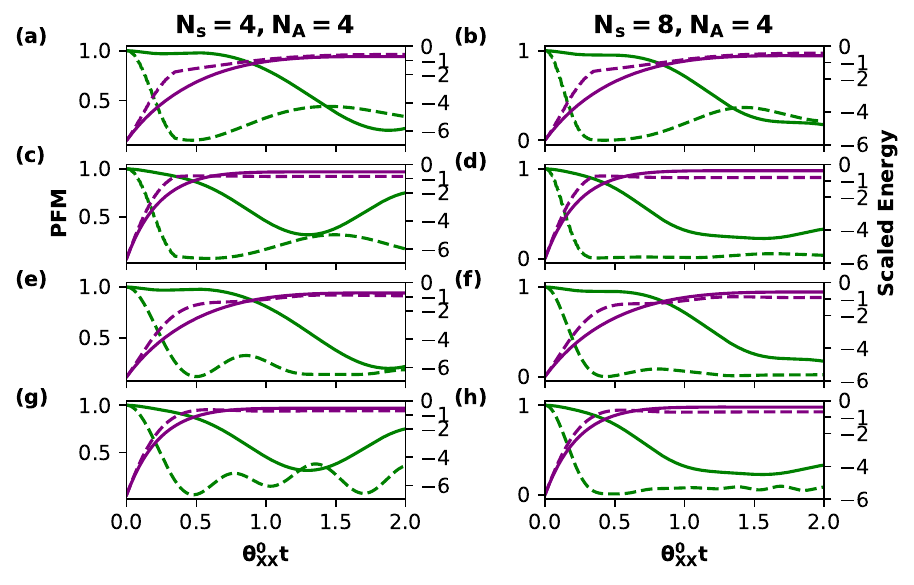}
    \caption{Infinite long-range TFIM in an open chain with $\theta_{XX}^{0} = 1.0,~\theta_{M}(0) = \theta_Z^{0} = 5\theta_{XX}^{0}$: time dependence of the projected probability PFM in the ground state of the mixer Hamiltonian $|1\rangle^{\otimes N_S}$(in green color with values given by the left side ticks) and the scaled energy (in purple with values given by the right side ticks) given by the ratio between the reduced system energy and the true system ground state energy from the imaginary time evolution from the always-on system Hamiltonian $H_{P}^{S}$. 
    Floquet kicks with the same kick angle $\theta = 0.004$ and the kick time interval $\Delta t_{K}$  between kicks is related to the Ising interaction $\theta_{XX}^{0}\Delta t_{K} = 0.0008/(0.5N_S)$.
    Floquet kicks: (a) $\theta_{Z}^{0}\tau = 0.8^{-1}$, (b) $\theta_{Z}^{0}\tau = 0.8^{-1}$, (c) $\theta_{Z}^{0}\tau = 1.6^{-1}$, (d) $\theta_{Z}^{0}\tau = 1.6^{-1}$. Continuous weak kicks with the constant kick angle $\theta \rightarrow 0, N_{K}\rightarrow \infty$: (e) $\theta_{Z}^{0}\tau = 0.8^{-1}$, (f) $\theta_{Z}^{0}\tau = 0.8^{-1}$, (g) $\theta_{Z}^{0}\tau = 1.6^{-1}$, (h) $\theta_{Z}^{0}\tau = 1.6^{-1}$.}
    \label{fig:FIG3}
\end{figure}

In Fig.~\ref{fig:FIG3} for ILR-TFM, the kick energy landscape versus the kick angle $\theta$ is insensitive to the connectivity of the TFIM since the reduced system energy is predominant by the one-local mixer Hamiltonian $ H^{S}_{M}(t\approx 0)$.
For the discrete kicks with slower and faster ramping, in subplots~\ref{fig:FIG3}(a-b) and~\ref{fig:FIG3}(c-d) respectively, we observe
similar behaviors, except that the typical annealing in purple manifests far more excitations. However, the quantum kicks are more efficient in
suppressing errors in the ILR-TFIM over NN-TFIM.
The same situation applies to the continuous kicks in subplots \ref{fig:FIG3}(e-h).

\begin{figure}[h]
    \centering
    \includegraphics[scale=0.65]{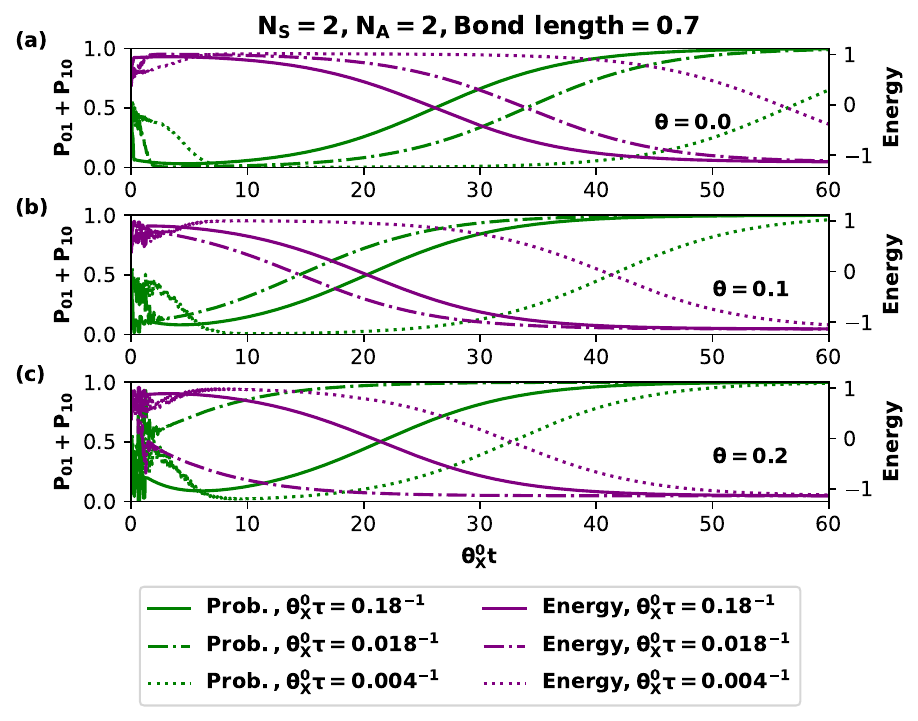}
    \caption{Time dependence of probability in odd parity subspace $|01\rangle$ and $|10\rangle$ (in green) and $\rm{H_2}$ system binding energy (in purple) with $N_{S} = 2, N_{A} = 2$, bond length $= 0.7$, and $\theta_{X}^{0} = 1.5$. The annealing rates $\theta_{X}^{0}\tau$ are shown in the attached legend. The cases without the kicks 
    are labeled by the kick angle $\theta = 0$ in the top subplot (a). The probability values (in green) are read by the left-handed ticks
    and the energy values (in purple) are read by the right-handed ticks.}
    \label{fig:FIG7}
\end{figure}

\begin{figure}[h]
    \centering
    \includegraphics[scale=0.65]{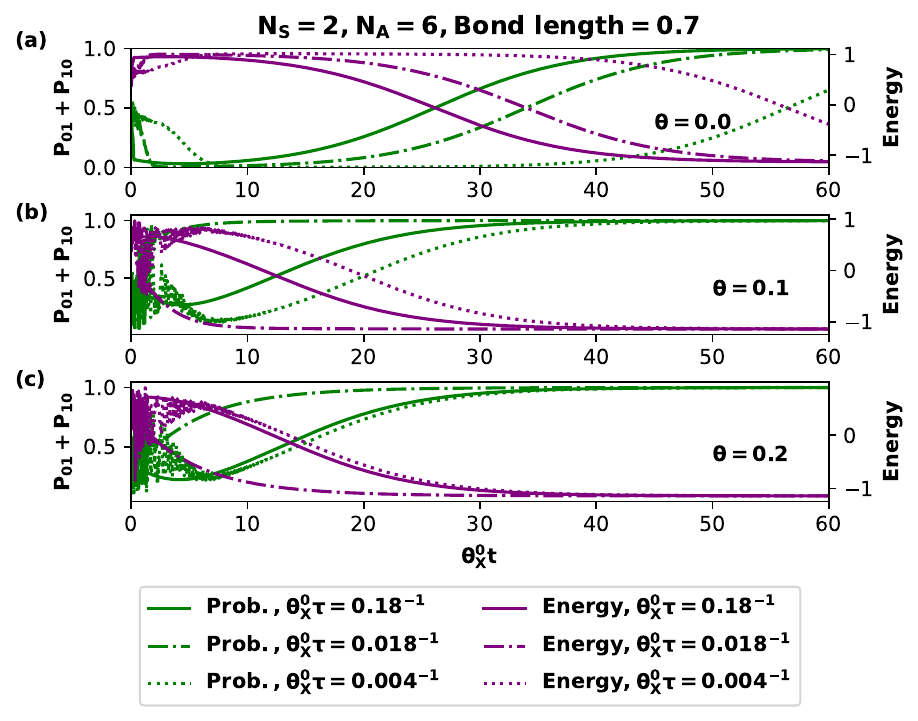}
    \caption{Time dependence of probability in odd parity subspace $|01\rangle$ and $|10\rangle$ (in green) and $\rm{H_2}$ system binding energy labeled by Energy (in purple) with $N_{S} = 2, N_{A} = 6$, bond length $= 0.7$, and $\theta_{X}^{0} = 1.5$. The annealing rates $\theta_{x}^{0}\tau$ are shown in the attached legend. The cases without the kicks 
    are labeled by the kick angle $\theta = 0$ in the top subplot (a). The probability values (in green) are read by the left-handed ticks
    and the energy values (in purple) are read by the right-handed ticks.}
    \label{fig:FIG8}
\end{figure}

In Fig.~\ref{fig:FIG7} with $N_A = 2$ and Fig.~\ref{fig:FIG8} with $N_A = 6$ with hydrogen bond length $=0.7~a.u.$,
we also observe the convergence to the true binding energy for any annealing rate. Still, with the clear advantages for the kicks, we show the speed convergence dependence is not monotonic concerning the annealing rates due to the complicated interplay of interactions with a comparable magnitude as shown in Table~\ref{TABLE:1}.

\end{document}